\documentclass[fp,twocolumn]{jpsj3}
\usepackage{txfonts}
\usepackage{color}
\usepackage{graphicx}

\title{Strong Bilayer Coupling Induced by the Symmetry Breaking in the Monoclinic Phase of BiS$_2$-Based Superconductors}

\author{Masayuki Ochi$^1$\thanks{E-mail: ochi@phys.sci.osaka-u.ac.jp}, Ryosuke Akashi$^2$, and Kazuhiko Kuroki$^1$}
\inst{$^1$Department of Physics, Osaka University, Machikaneyama-cho, Toyonaka, Osaka 560-0043, Japan \\
$^2$Department of Physics, The University of Tokyo, Hongo, Bunkyo-ku, Tokyo 113-0033, Japan} 

\abst{We perform first-principles band structure calculations for the tetragonal and monoclinic structures of LaO$_{0.5}$F$_{0.5}$BiS$_2$.
We find that the Bi $6p_{x,y}$ bands on two BiS$_2$ layers exhibit a sizable splitting at the X = ($\pi$, 0, 0) and several other \textbf{k}-points for the monoclinic structure.
We show that this feature originates from the inter-BiS$_2$ layer coupling strongly enhanced by the symmetry breaking of the crystal structure.
The Fermi surface also shows a large splitting and becomes anisotropic with respect to the $k_x$ and $k_y$ directions in the monoclinic structure, whereas it remains almost flat with respect to the $k_z$ direction.}

\begin{document}
\maketitle

\section{Introduction}
Superconductivity has been a central subject in condensed matter physics both from the theoretical viewpoint as a fertile playground for various physics and from the applicational viewpoint as a resource of innovative devices.
One of the most well-known and prominent classes of superconductors is cuprates~\cite{Cuprate}, where copper and oxygen atoms form superconducting square layers.
The discovery of BiS$_2$-based superconductors~\cite{BiS2,BiS2_2,BiS2review,BiS2review2,BiS2review3} attracted much attention owing to their similarity to layered superconductors such as cuprates and iron pnictides~\cite{iron1,iron2}: BiS$_2$-based superconductors have superconducting square layers consisting of bismuth and sulfur atoms, the in-plane $p_{x,y}$ orbitals of which constitute the Fermi surface~\cite{Usui} with a strong two-dimensionality owing to the existence of blocking layers.
Its pairing mechanism is still under debate~\cite{BiS2review,BiS2review2,BiS2review3}, and active investigation, e.g., on what determines the critical temperature $T_c$, is in progress.

It has recently been reported that an abrupt increase (more than double in many cases) in $T_c$ of $R$O$_{0.5}$F$_{0.5}$BiS$_2$ ($R=$ La, Ce, Pr, Nd) and several related compounds~\cite{PT1,PT2,PTstrct,PT3,PT4,PT5,PT6,PT7,PT8,PT9} coincides with the structural phase transition from a $P$4/$nmm$ tetragonal structure to a $P$2$_1$/$m$ monoclinic structure upon pressure application~\cite{PTstrct,PT7}.
Because these studies suggest that a structural change is a key factor for the increase in $T_c$, an investigation of the monoclinic structure should give us an important clue to understanding the pairing mechanism in BiS$_2$-based superconductors. However, most theoretical studies have focused on the tetragonal structure~\cite{note_Martins}.
It is also noteworthy that a recent X-ray diffraction experiment reported that the structure of LaOBiS$_2$ single crystal belongs to a $P$2$_1$/$m$ space group at ambient pressure~\cite{SymLow}.

In this study, we perform first-principles band structure calculations for the tetragonal and monoclinic structures of LaO$_{0.5}$F$_{0.5}$BiS$_2$.
The Bi $6p_{x,y}$ bands on two BiS$_2$ layers, which constitute the lowest conduction bands near the X point, are found to exhibit a large splitting for the monoclinic structure.
Because the superconducting transition is induced by electron doping into these bands, their large splitting should considerably affect the superconductivity.
This motivates us to scrutinize their band splitting under several conditions and reveal its mechanism.
For this purpose, we construct a tight-binding model consisting of the Bi $6p_{x,y}$ and S $3p_{x,y}$ orbitals for LaOBiS$_2$.
Looking into the obtained Bloch states, we find that the high symmetry of the tetragonal structure guarantees cancellation among some inter-BiS$_2$ layer hopping paths, which holds only partially in the monoclinic one.
A strong two-dimensionality of the Fermi surface is almost retained while the anisotropy with respect to the $x$- and $y$-directions emerges in the monoclinic structure.
We also evaluate the effect of the substitution of oxygen atoms by fluorine atoms, which is usually carried out for electron doping, and that of the spin-orbit coupling (SOC) on the band splitting.
Our findings are robust against a change of atomic species and so serve as important knowledge for understanding the pairing mechanism of the BiS$_2$-based superconductors.

\section{Computational Conditions\label{2}}
\begin{figure}
\includegraphics[width=8.5cm]{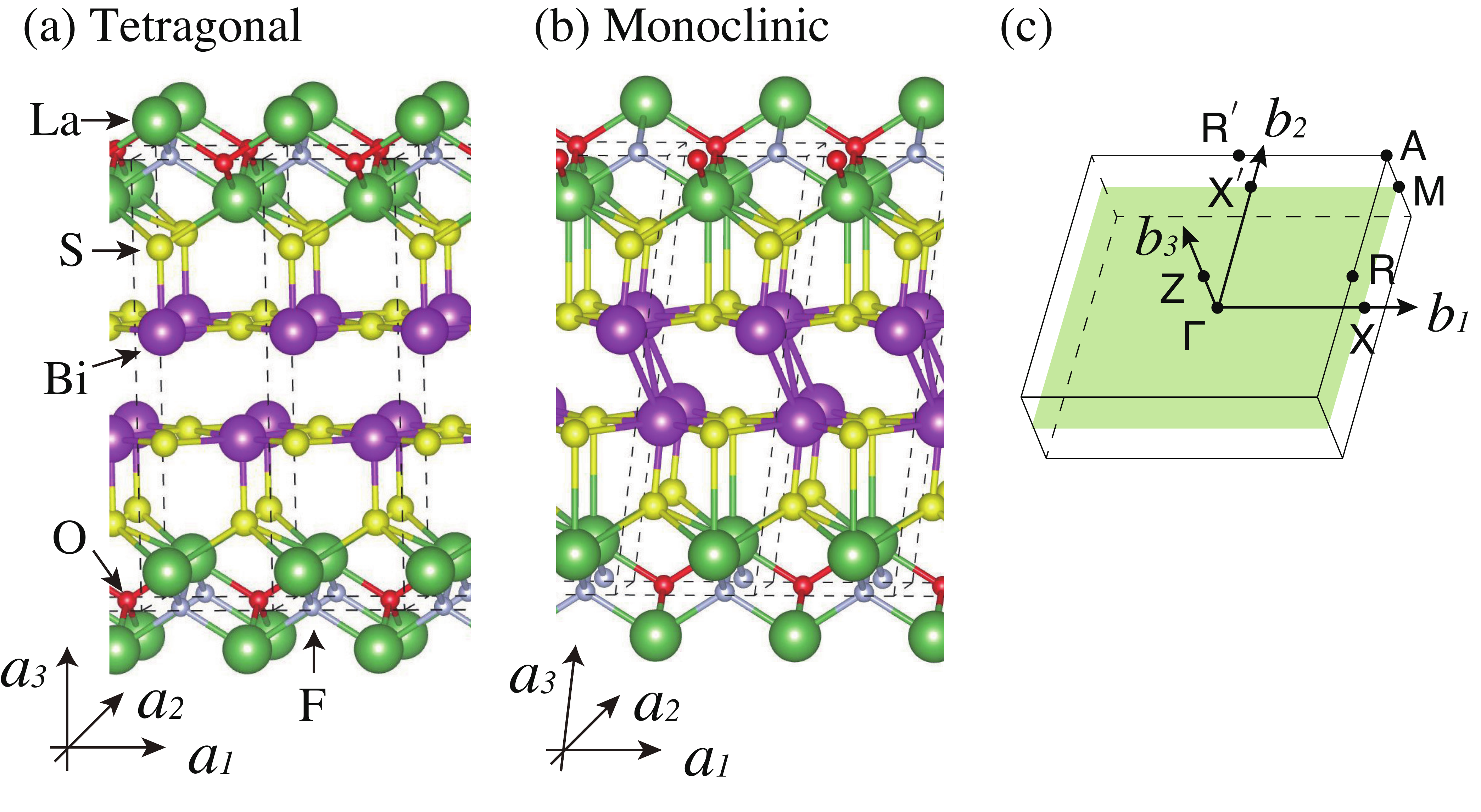}
 \caption{(Color online) (a) Tetragonal and (b) monoclinic crystal structures of LaO$_{0.5}$F$_{0.5}$BiS$_2$ drawn using VESTA software~\cite{VESTA}. (c) Definition of \textbf{k}-points.}
\label{fig:crystal}
\end{figure}

For first-principles band structure calculations, we used the Perdew-Burke-Ernzerhof exchange-correlation functional~\cite{PBE} and the full-potential linearized augmented plane-wave method as implemented in the \textsc{wien2k} code~\cite{wien2k}.
Crystal structures of LaO$_{0.5}$F$_{0.5}$BiS$_2$ were taken from Refs.~\citen{APstrct} ($a=4.07063$ \AA, $c=13.4848$ \AA) and \citen{PTstrct} ($a=4.042$ \AA, $b=4.059$ \AA, $c=12.809$ \AA, $\beta$=97.31$^\circ$), and are shown in Fig.~\ref{fig:crystal}(a)-(b) for the tetragonal and monoclinic structures, respectively.
Here, we replaced one oxygen atom with one fluorine atom in the unit cell to represent electron doping, as shown in the figures.
Without such replacement, the tetragonal and monoclinic structures belong to the space groups $P$4/$nmm$ and $P$2$_1$/$m$, respectively.
The muffin-tin radii for La, O, F, Bi, and S atoms, $r_{\rm La}$, $r_{\rm O}$, $r_{\rm F}$, $r_{\rm Bi}$, and $r_{\rm S}$, were set to 2.35, 2.07, 2.12, 2.50, and 2.01 Bohr, respectively.
The maximum modulus for the reciprocal lattice vectors $K_{\rm max}$ was chosen so that $r_{\rm S} K_{\rm max}$ =7.00.

\section{Results and discussion\label{3}}
\subsection{Band structures and Fermi surfaces}
\begin{figure*}
\includegraphics[width=18cm]{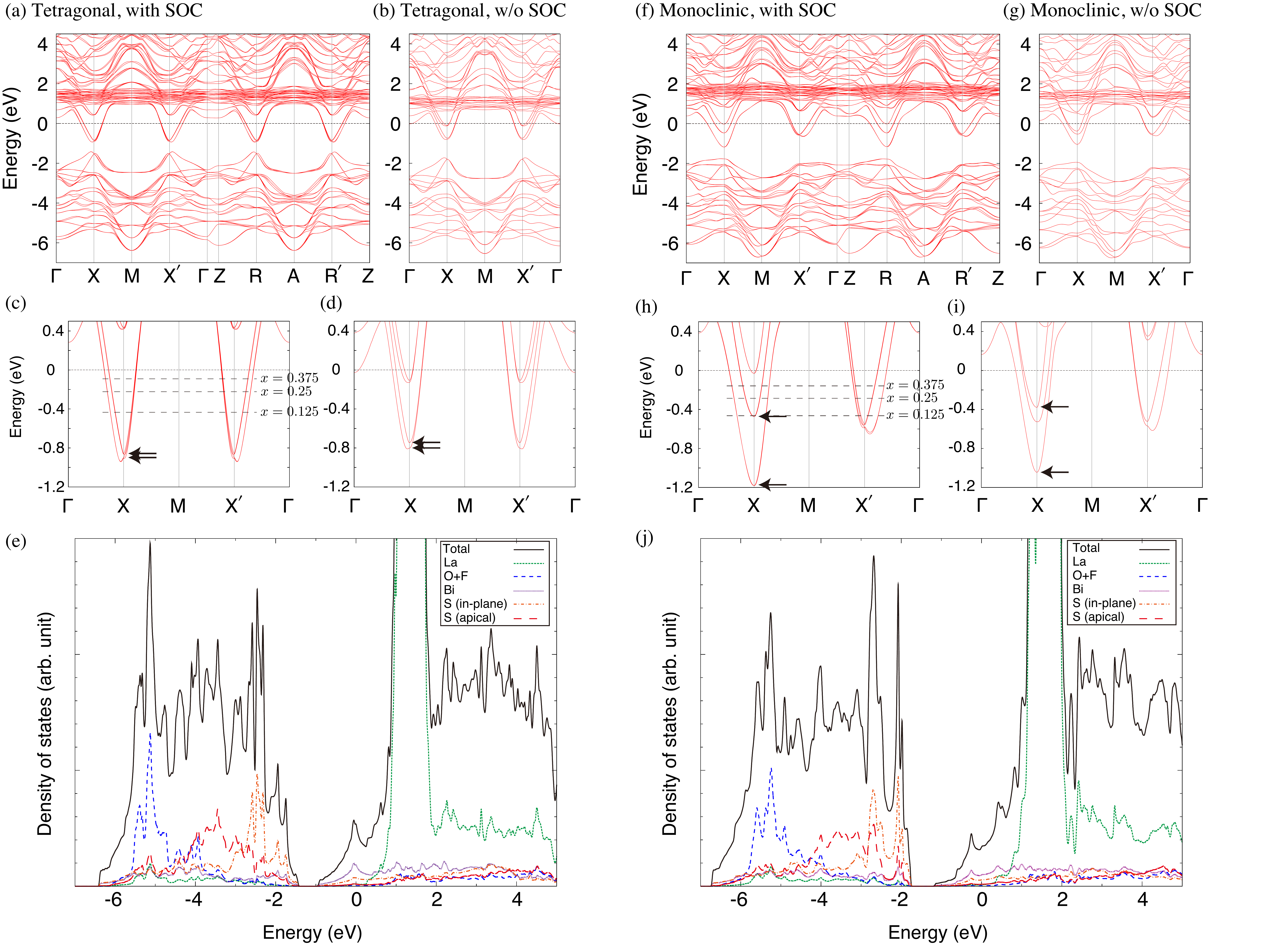}
 \caption{(Color online) Calculated band structures for LaO$_{0.5}$F$_{0.5}$BiS$_2$ of the tetragonal structure (a) with and (b) without SOC, and (f)-(g) those for the monoclinic one. The enlarged figures near the Fermi level along the $\Gamma$-X-M-X$'$-$\Gamma$ line for (a), (b), (f), and (g) are shown in (c), (d), (h), and (i), respectively. Black arrows in panels (c), (d), (h), and (i) show the Bi $p_x$ states on two BiS$_2$ layers at the X point.
Dashed lines in panels (c) and (h) indicate the Fermi levels corresponding to several amounts of electron doping as represented by LaO$_{1-x}$F$_x$BiS$_2$. For example, 0 eV trivially corresponds to $x=0.5$. The calculated (partial) density of states for the tetragonal and monoclinic structures with SOC are shown in (e) and (j), respectively. In panels (e) and (j), S (in-plane) denotes the S atoms located in the square lattice consisting of Bi and S atoms, while S (apical) denotes S atoms outside the square lattice. The Fermi level is set to 0 eV for each panel.}
\label{fig:band}
\end{figure*}

Figure~\ref{fig:band}(a)-(j) presents the calculated band structures and (partial) density of states for the tetragonal and monoclinic structures with/without an inclusion of the SOC, where the definition of the \textbf{k}-points used in this paper is shown in Fig.~\ref{fig:crystal}(c).
In the tetragonal structure, the X and X$'$ points as well as the R and R$'$ points are equivalent.
The overall band structures are similar between the two crystal structures, but we find a large splitting for the conduction band bottom at the X=($\pi$, 0, 0) point in the monoclinic structure regardless of the inclusion of SOC.
As analyzed in the previous study~\cite{Usui} and also shown here [as indicated by black arrows in Fig.~\ref{fig:band}(c)(d)(h)(i)], the conduction band bottoms at the X point correspond to the Bloch states on two BiS$_2$ layers, and hence the feature observed here should be related to the bilayer coupling between them.
Note that the complete degeneracy of the conduction band bottom at the X point for the tetragonal LaOBiS$_2$ is slightly lifted by the substitution of O atoms with F atoms here, which shall be analyzed later in this paper.
It is also worth noting that the large band splitting is not observed at the X$'$=(0, $\pi$, 0) point in the monoclinic structure.

\begin{figure}
 \includegraphics[width=8.5cm]{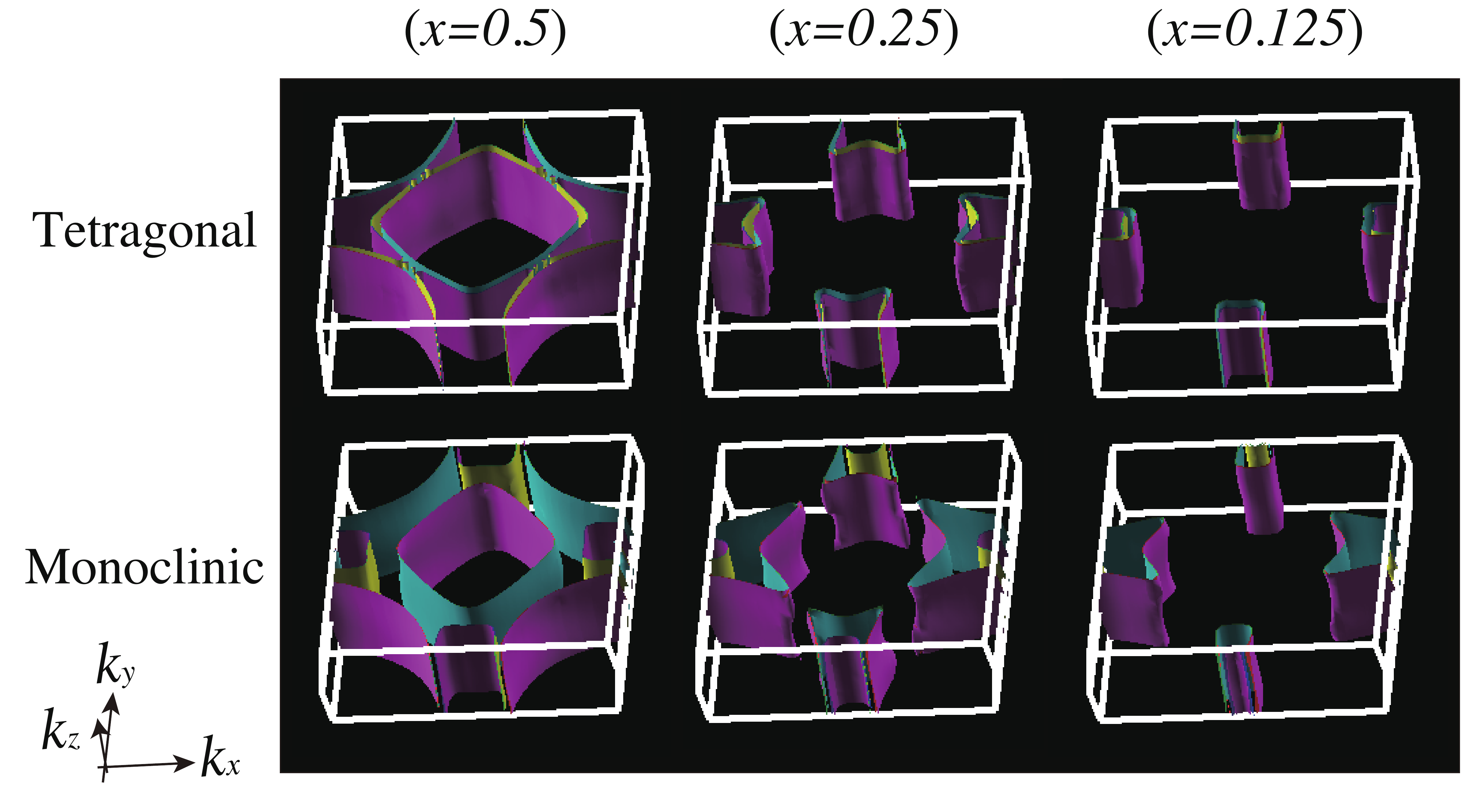}
 \caption{(Color online) Fermi surfaces of the tetragonal and monoclinic LaO$_{1-x}$F$_{x}$BiS$_2$ with an inclusion of SOC drawn using XCrySDen software~\cite{XCrySDen}.}
 \label{fig:FS}
\end{figure}

We find that the band splitting induces a noticeable change in the Fermi surface topology, as shown in  Fig.~\ref{fig:FS}, which was calculated with an inclusion of SOC.
The rigid band approximation was employed for electron doping in LaO$_{1-x}$F$_x$BiS$_2$, namely, the band structure was fixed to that of LaO$_{0.5}$F$_{0.5}$BiS$_2$.
Here, the band splitting seen above makes the nearly degenerate Fermi surfaces in the tetragonal structure much separated in the monoclinic one. For example, for $x=0.25$ doping in the monoclinic structure, the small Fermi pockets near the X point almost disappear.
The inequivalency between the X and X$'$ points in the monoclinic structure yields a strong anisotropy of the Fermi surface for a small amount of electron doping, whereas it is accidentally inconspicuous for $x=0.5$ doping.
The strong two-dimensionality of the Fermi surface is retained even in the symmetry-broken monoclinic structure, which means that the electronic states on the BiS$_2$ planes in the respective unit cells are well decoupled by the existence of the blocking layer consisting of La, O, and F atoms.

\subsection{Analysis using a tight-binding model}

\subsubsection{Setup}

\begin{figure}
\begin{center}
 \includegraphics[width=7cm]{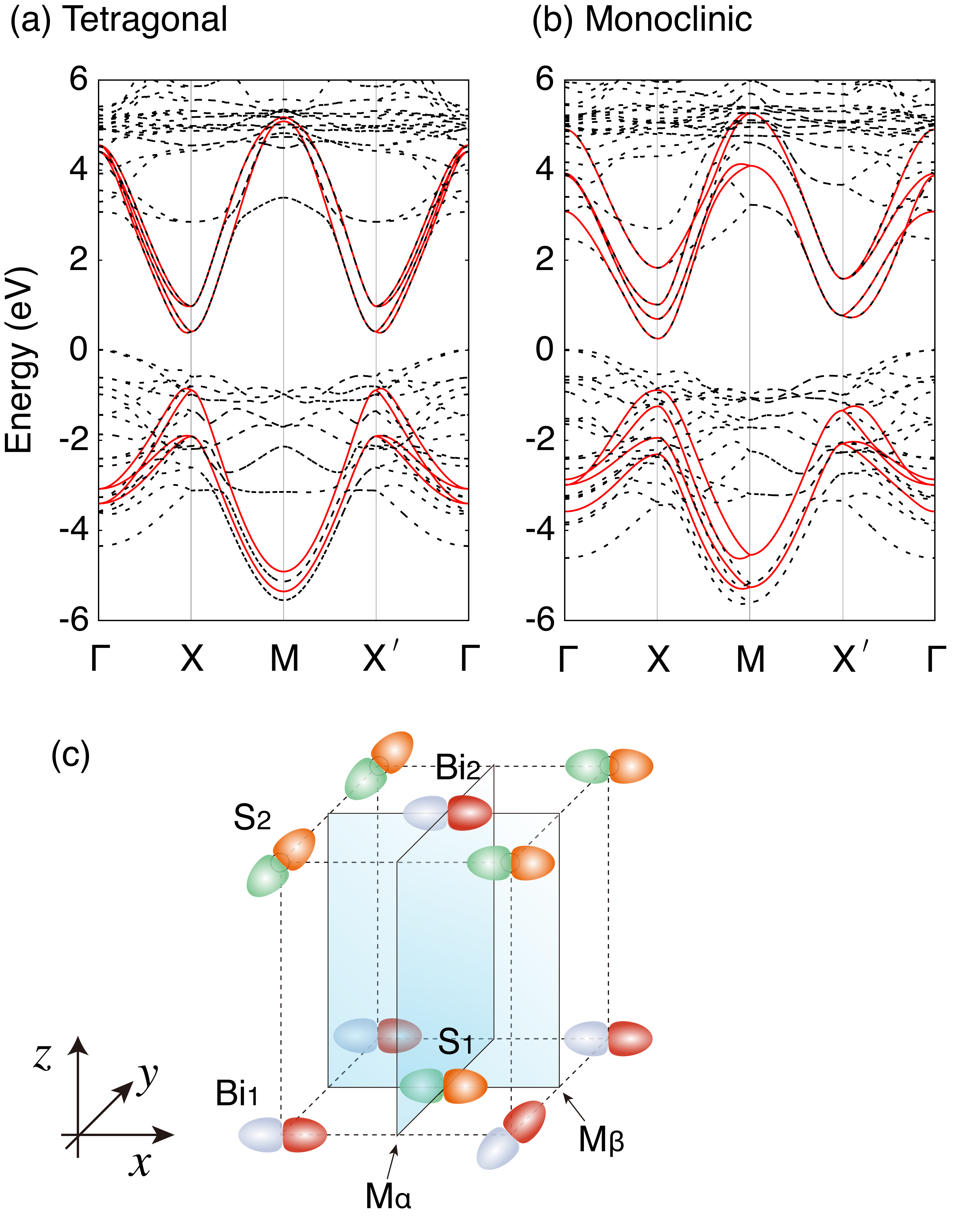}
 \caption{(Color online) Band structures of the tight-binding model (red solid lines) together with those obtained in first-principles calculations (black broken lines) for the (a) tetragonal and (b) monoclinic structures of LaOBiS$_2$ without an inclusion of SOC. The valence band top is set to 0 eV. (c) Notation of atomic sites and mirror planes in our bilayer tight-binding model for the tetragonal structure.}
 \label{fig:model}
\end{center}
\end{figure}

We have seen that the splitting of the conduction band bottom at the X=($\pi$, 0, 0) point is a key difference between the two crystal structures. 
To analyze the origin of this feature, we constructed a tight-binding model consisting of Bi $6p_{x,y}$ and S $3p_{x,y}$ orbitals for the tetragonal and monoclinic structures of the parent compound LaOBiS$_2$.
For the tetragonal structure, we used an experimental crystal structure taken from Ref.~\citen{TetraStruct} ($a=4.05$ \AA, $c=13.74$ \AA). For the monoclinic structure, we used the same crystal structure as LaO$_{0.5}$F$_{0.5}$BiS$_2$, where the F atom is virtually replaced with the O atom.
We employed the maximally localized Wannier functions~\cite{Wannier1,Wannier2,Wannier90,Wien2Wannier} constructed from the first-principles band structures. We neglected SOC and omitted the spin index here for simplicity. 
The effect of SOC on the band structure shall be analyzed later in this paper.
Band structures of our tight-binding models together with the ab initio ones are presented in Fig.~\ref{fig:model}(a)-(b). 
Our main purpose is to investigate the electronic structures near the conduction band bottoms at the X and X$'$ points, which dominate the electron-doped superconductivity. Hence, we adopt the present model that accurately reproduces the band structure in those regions. To investigate the change of the Bloch states near the X(X$'$) point, we shall examine the Bloch states not only at the X and X$'$ points but also along the $\Gamma$-X(X$'$) and X(X$'$)-M lines.
As noted in the previous section, a minor effect of the replacement of F atoms with O atoms will be addressed later in this paper.

\begin{figure}
  \includegraphics[width=8.5cm]{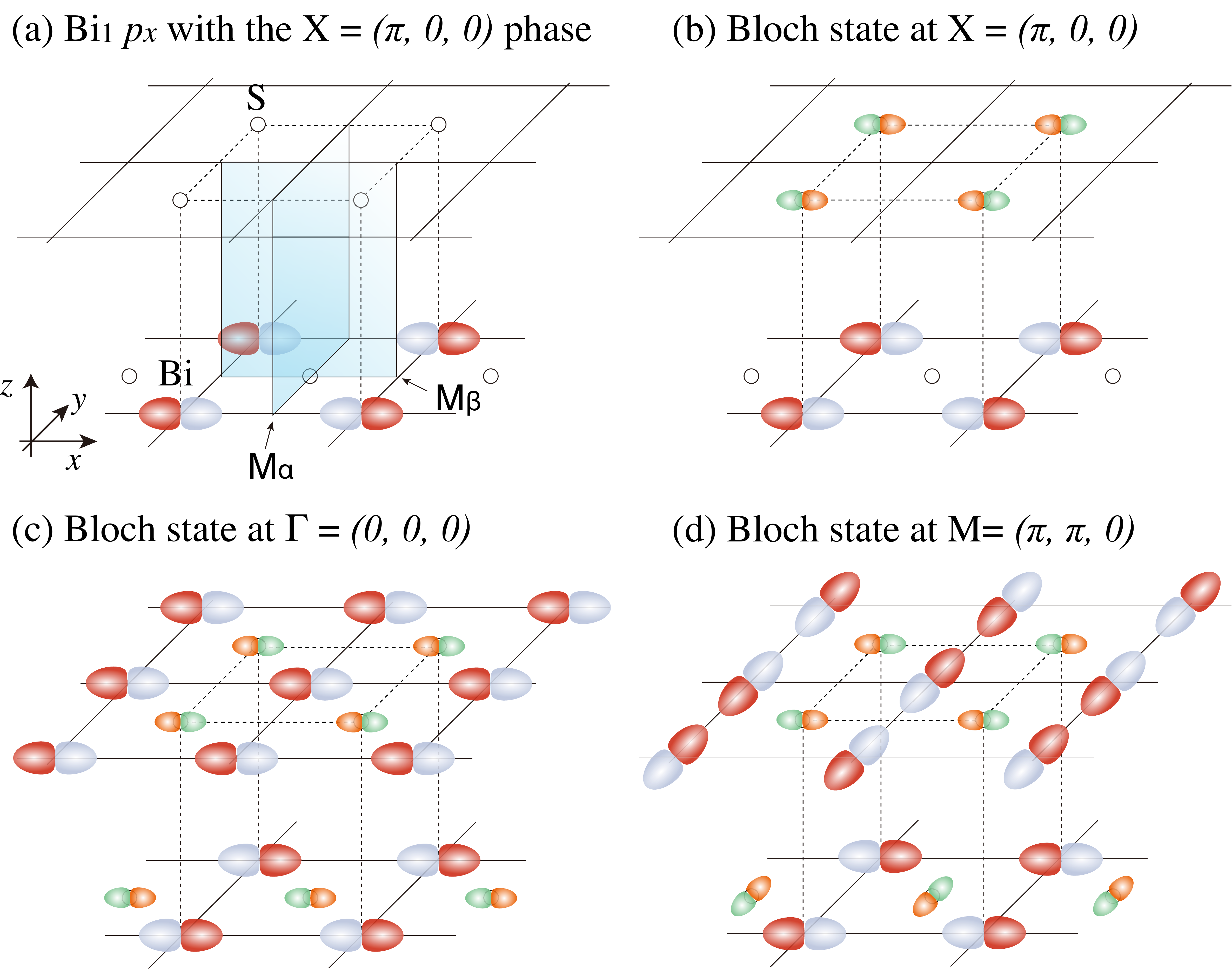}
 \caption{(Color online) Schematic pictures of (a) Bi$_1$ $p_x$ orbital with the X phase, and of the Bloch states in the lowest conduction bands at the (b) X, (c) $\Gamma$, and (d) M points for the tetragonal structure.}
 \label{fig:orbitals}
\end{figure}

We mainly focus on the tight-binding model for the tetragonal structure.
Its site indices and mirror planes used in later discussion are shown in Fig.~\ref{fig:model}(c).
With this setting, the reflection symmetry with respect to the M$_{\alpha}$ plane is broken by the structural transition to the monoclinic structure.
The present model consists of eight atomic orbitals in each unit cell: Bi$_{1,2}$ $p_{x,y}$ and S$_{1,2}$ $p_{x,y}$, where 1 and 2 are the layer indices~\cite{foot:pXY}.
We define an `atomic orbital with the Bloch phase $\mathbf{k}$ = ($k_x$, $k_y$, $k_z$)' as the linear combination of atomic orbitals on each site: $\phi_{i,\mathbf{k}}=\sum_{\mathbf{R}} \mathrm{exp}[\mathrm{i}\mathbf{k}\cdot\mathbf{R}] \phi_{i,\mathbf{r}_i+\mathbf{R}}$, where $\mathbf{r}_i$ and $\mathbf{R}$ are the site coordinate and lattice vector, respectively, with $\phi_{i,\mathbf{r}_i+\mathbf{R}}$ being an atomic orbital $i$ on the site $\mathbf{r}_i+\mathbf{R}$.
Such a linear combination forms a basis set to represent Bloch states at a specific \textbf{k}-point.
Figure~\ref{fig:orbitals}(a) shows an example: a Bi$_1$ $p_x$ orbital with the X=($\pi$, 0, 0) phase.
Table~\ref{table:orb} shows the parities of the atomic orbitals with some Bloch phases for the tetragonal structure, which are used for the following discussion.

Our tight-binding Hamiltonian is invariant under an exchange of the layer index (e.g., Bi$_1$ and Bi$_2$) for both tetragonal and monoclinic structures.
Such equivalency between two BiS$_2$ layers is guaranteed by the inversion and time-reversal symmetries in the crystal structure of LaOBiS$_2$ as follows: the former operation exchanges the two layers and transforms the crystal momentum $\bf{k}$ into $-\bf{k}$, and then the latter one restores the sign of $\bf{k}$.
From the next section, we shall examine how the atomic orbitals on the two equivalent layers are coupled at each \textbf{k}-point.
For the tetragonal structure, the following argument is valid even if the time-reversal symmetry is broken, such as in the ferromagnetic CeOBiS$_2$~\cite{Ce}. This is because the glide reflection ($a_1$, $a_2$, $a_3$) $\to$ ($a_1+0.5$, $a_2+0.5$, $-a_3$) instead ensures the layer equivalency for the Bloch states on the $k_z=0$ plane.

\begin{table*}
\begin{center}
\begin{tabular}{c c c c c}
\hline
\hline
 M$_{\alpha}$ &\multicolumn{2}{c}{+1} & \multicolumn{2}{c}{-1} \\
 M$_{\beta}$ & +1 & -1 & +1 & -1 \\
\hline
$\Gamma$ & - &  (Bi$_1$ $p_y$, S$_2$ $p_y$), (Bi$_2$ $p_y$, S$_1$ $p_y$) & (Bi$_1$ $p_x$, S$_2$ $p_x$), (Bi$_2$ $p_x$, S$_1$ $p_x$) & - \\
X & (Bi$_1$ $p_x$, S$_2$ $p_x$) & (Bi$_2$ $p_y$, S$_1$ $p_y$) & (Bi$_2$ $p_x$, S$_1$ $p_x$) & (Bi$_1$ $p_y$, S$_2$ $p_y$)\\
X$'$ & (Bi$_1$ $p_y$, S$_2$ $p_y$) & (Bi$_2$ $p_y$, S$_1$ $p_y$) & (Bi$_2$ $p_x$, S$_1$ $p_x$) & (Bi$_1$ $p_x$, S$_2$ $p_x$)\\
M & - & (Bi$_1$ $p_x$, S$_2$ $p_x$), (Bi$_2$ $p_y$, S$_1$ $p_y$) & (Bi$_1$ $p_y$, S$_2$ $p_y$), (Bi$_2$ $p_x$, S$_1$ $p_x$) & -\\
\hline
\hline
\end{tabular}
\end{center}
\caption{\label{table:orb} Parities of the atomic orbitals with four Bloch phases for the reflection operator with respect to the two mirror planes M$_{\alpha}$ and M$_{\beta}$ defined in Fig.~\ref{fig:model}(c), for the tetragonal structure. Some pairs of orbitals, which are located at the same $x$-, $y$- but different $z$-coordinates (e.g., Bi$_1$ $p_y$ and S$_2$ $p_y$), always have the same parity and are denoted by parentheses for visibility.}
\end{table*}

\subsubsection{X and X$'$ points}

First, we consider Bloch states at the X=($\pi$, 0, 0) point in the tetragonal structure.
Because of the mirror symmetries, Hamiltonian at the X point consists of four independent blocks with different parities, as shown in Table~\ref{table:orb}.
This block structure of Hamiltonian forbids the hybridization between two equivalent states that can be transformed into each other by the layer exchange, which allows a two-fold degeneracy.
For example, the band energy for the (M$_{\alpha}$, M$_{\beta}$) = (+1, +1) block where the Bloch states consist of Bi$_1$ $p_x$ and S$_2$ $p_x$, should be the same as that for the (M$_{\alpha}$, M$_{\beta}$) = ($-$1, +1) block where the Bloch states consist of Bi$_2$ $p_x$ and S$_1$ $p_x$.
One of the Bloch states at the conduction band bottom is shown in Fig.~\ref{fig:orbitals}(b). Because the difference between the onsite energies of Bi and S $p_{x,y}$ (about 2.0 eV) is much larger than the coupling between Bi $p_x$ and S $p_x$ on the neighboring layers (0.18 eV), the Bloch state shown in this figure is the Bi $p_x$ orbitals on one layer slightly hybridized with the S $p_x$ orbitals on the neighboring layer.
Here, we have seen that the four Bi orbitals on the two planes, namely, Bi$_{1,2}$ $p_{x,y}$, are completely decoupled at the X point in the tetragonal structure.
Note that the X and X$'$ points are equivalent in the tetragonal structure.

However, in the monoclinic structure, the symmetries used in the above discussion are partially broken. As mentioned earlier, the symmetry for the layer exchange and the mirror symmetry with respect to the M$_{\beta}$ plane remain, but that for the M$_{\alpha}$ plane is lost. Therefore, at the X point, Hamiltonian consists of two blocks: four $p_x$ orbitals constitute one block, and four $p_y$ orbitals constitute the other one (see Table~\ref{table:orb} and only consider the M$_{\beta}$ parity).
Here, the equivalent states with respect to the layer exchange always belong to the same block, which allows the hybridization between them and therefore lifts the degeneracy.
On the other hand, the two-fold degeneracy at the X$'$ point is retained.
This is because, at the X$'$ point, Hamiltonian consists of two blocks: $\{$Bi$_1$ $p_y$, Bi$_2$ $p_x$, S$_1$ $p_x$, S$_2$ $p_y$$\}$ and $\{$Bi$_1$ $p_x$, Bi$_2$ $p_y$, S$_1$ $p_y$, S$_2$ $p_x$$\}$, where a Bloch state in one block can be transformed to another Bloch state in the other block by layer exchange, which means that our discussion presented in the previous paragraph can be applied to this case.
Such inequivalency between the X and X$'$ points is the origin of the large anisotropy of the Fermi surface that we have seen before in this paper.

\subsubsection{$\Gamma$-X(X$'$) line}

Next, we shall look at the $\Gamma$ point. As seen in Table~\ref{table:orb}, the Bloch state at the $\Gamma$ point is a mixture of many orbitals on both planes even in the tetragonal structure. 
One of the Bloch states in the lowest conduction bands at the $\Gamma$ point in the tetragonal structure is shown in Fig.~\ref{fig:orbitals}(c).
However, owing to the equivalency between the $x$- and $y$-directions, a two-fold degeneracy is present for the tetragonal structure.
In other words, one Bloch state in the (M$_{\alpha}$, M$_{\beta}$) = (+1, $-$1) block consisting of $p_y$ orbitals should have an equivalent counterpart in the (M$_{\alpha}$, M$_{\beta}$) = ($-$1, +1) block consisting of $p_x$ orbitals.
This degeneracy is obviously lifted for the monoclinic structure, and also on the $\Gamma$-X(X$'$) line except at the endpoints for both crystal structures since $k_x\neq k_y$.

\subsubsection{X(X$'$)-M line}

Here, we consider the X(X$'$)-M line for the tetragonal structure. Along this line, only the M$_{\alpha}$(M$_{\beta}$) parity can be defined for atomic orbitals, and then the Hamiltonian consists of two blocks (see the M point in Table~\ref{table:orb}). 
Similarly to the X(X$'$) point, the symmetry for layer exchange guarantees a two-fold degeneracy along this line.
In the monoclinic structure, where the M$_{\alpha}$ symmetry is broken, such degeneracy remains only on the X$'$-M line.
One of the Bloch states in the lowest conduction bands at the M point in the tetragonal structure is shown in Fig.~\ref{fig:orbitals}(d).

\subsubsection{Degeneracy for non-symmorphic space groups}

In this section, we shall return to the first-principles band structure and the whole crystal structure including the blocking layers.
Non-symmorphic space groups of the crystal structure always exhibit the band degeneracy at some \textbf{k}-points~\cite{Volker,group1,group2,group3}, which is utilized, for example, to unfold the Brillouin zone using the glide reflection symmetry in iron-based superconductors (see, e.g., Ref.~\citen{GlideFe}).
As a matter of fact, the above-mentioned degeneracy can also be explained by an argument based on the non-symmorphic space group.
In our case, a key symmetry operation is a two-fold screw rotation $S$: ($a_1$, $a_2$, $a_3$) $\to$ ($-a_1$, $a_2+0.5$, $-a_3$) for both monoclinic and tetragonal structures of LaOBiS$_2$ [see Fig.~\ref{fig:crystal}(b), but note that the substitution of O with F atom breaks this symmetry].
Using this symmetry and the time-reversal symmetry, the degeneracy on the X$'$-M line for both structures can be proven as follows: first, we consider the symmetry operation $ST$, where $T$ is the time-reversal operator.
Here, $ST$ does not change the crystal momentum $\bf{k}$ on the X$'$-M line where $k_y=\pi$.
Next, we assume that the Bloch state $\phi_{\bf{k}}$ on the X$'$-M line is not degenerate, which requires this state to be an eigenstate of $ST$: $ST\phi_{\bf{k}}=\alpha\phi_{\bf{k}}$ with a complex eigenvalue $\alpha$.
By operating $ST$ again on this state, we get $STST\phi_{\bf{k}}=\alpha\alpha^*\phi_{\bf{k}}$.
On the other hand, because $STST$ transforms a function $f(\bf{r})$ into $f(\bf{r}+$(0, 1, 0)$)$ by definition, $STST\phi_{\bf{k}}=-\phi_{\bf{k}}$ owing to the Bloch phase.
However, $|\alpha|^2=-1$ cannot be satisfied, which denies the assumption of non-degeneracy.

The above proof is valid regardless of the orbital character as seen in Fig.~\ref{fig:model}(a)-(b) where all the bands along the X$'$-M line exhibit the degeneracy. Hence, the degeneracy itself does not give us a physical insight that, for example, such degeneracy on the X(X$'$) point is related to the bilayer coupling in our system. This is why we have analyzed the tight-binding model and investigated which orbitals are coupled or decoupled at each \textbf{k}-point.

\subsubsection{Origin of the large bilayer splitting}

We have mainly focused on the presence or absence of the band degeneracy, but here we briefly discuss the size of the band splitting.
A bismuth atom on one BiS$_2$ plane is just above a sulfur atom on the other plane (i.e., their $x$- and $y$-coordinates are common) in the tetragonal structure, which is not the case with respect to the $x$-coordinate for the monoclinic one [see Fig.~\ref{fig:crystal}(a)-(b)].
This might be one reason why the c-axis length of the monoclinic cell is much smaller than that for the tetragonal one, which induces a large enhancement of the interlayer coupling and the resulting sizable band splitting.
In fact, the differences in the $z$-coordinates of two Bi atoms on the neighboring planes are 3.3 and 2.9 \AA\ for the tetragonal and monoclinic structures of LaO$_{0.5}$F$_{0.5}$BiS$_2$, respectively.
The importance of the shortening of the Bi-Bi distance (4.4 to 3.6 \AA\ in our setup) has been pointed out in the previous study~\cite{PTstrct}.
Actually, we find that direct interlayer hopping paths between Bi atoms have sizable amplitudes in monoclinic LaOBiS$_2$: 0.14 eV for Bi $p_x$-Bi $p_x$ and 0.27 eV for Bi $p_y$-Bi $p_y$, whereas those in the tetragonal structure are smaller: 0.09 eV for both hopping processes.
Other sizable interlayer hopping amplitudes are 0.11 eV for Bi $p_x$-S $p_x$ and 0.09 eV for Bi $p_y$-S $p_y$ in the monoclinic structure, whereas those in the tetragonal structure are 0.18 eV for both hopping processes.

\subsubsection{Effect of F substitution and SOC on the band structures}

\begin{figure}
\begin{center}
\includegraphics[width=8.5 cm]{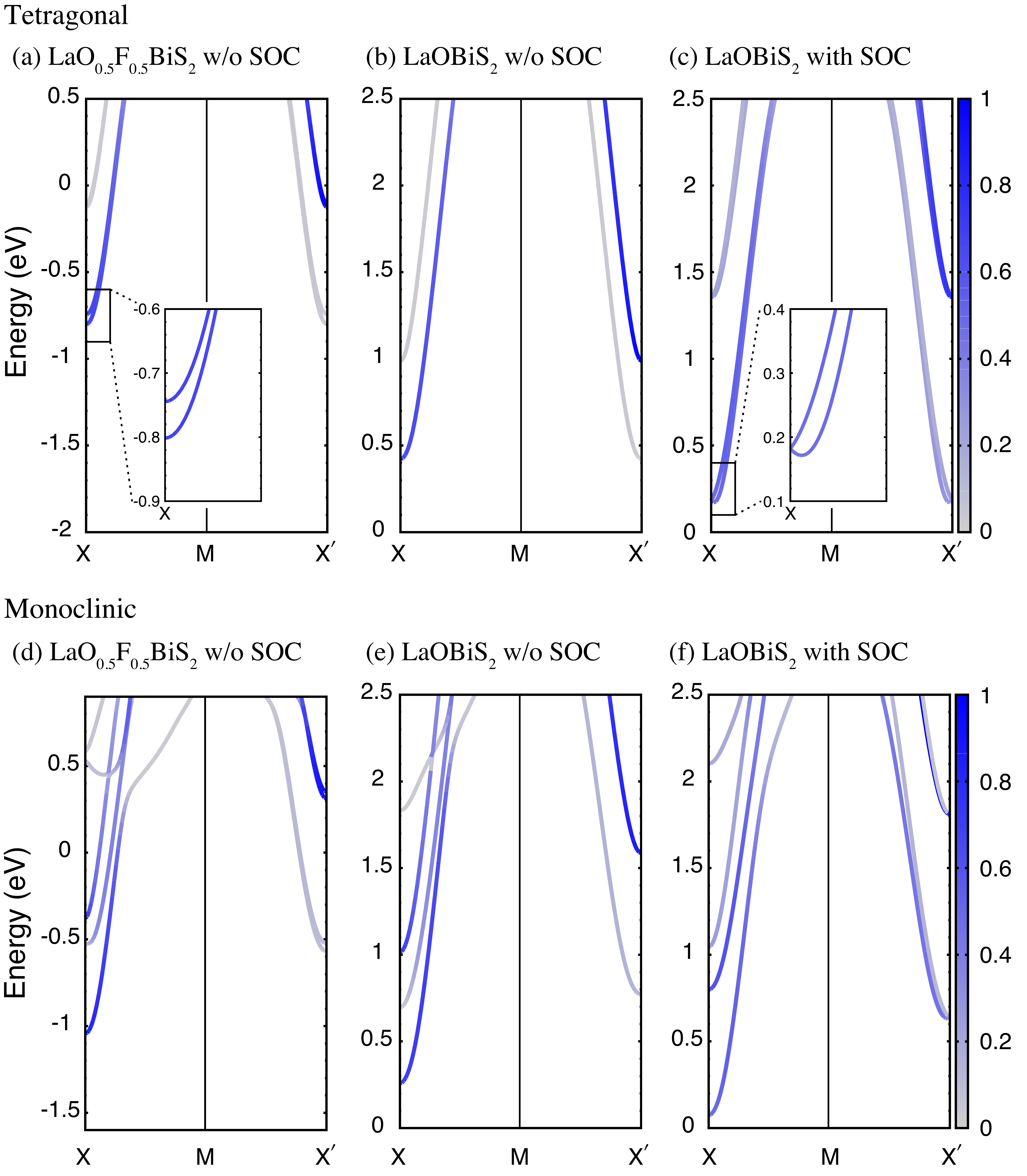}
 \caption{(Color online) First-principles band structures of (a) LaO$_{0.5}$F$_{0.5}$BiS$_2$ without SOC, (b) LaOBiS$_2$ without SOC, (c) LaOBiS$_2$ with SOC for the tetragonal structure, and (d)-(f) those for the monoclinic one. The color shows the Bi $p_x$ weight renormalized so that its sum over the eight lowest conduction bands including the spin degrees of freedom at the X$'$ point (i.e., all the bands shown here) is equal to 4. The Fermi level for LaO$_{0.5}$F$_{0.5}$BiS$_2$ and the valence top for LaOBiS$_2$ are set to 0 eV.}
\label{fig:band_closeup}
\end{center}
\end{figure}

Here, we comment on two issues neglected in our model analysis: substitution of O atoms with F atoms and SOC.
First, we focus on the effect of F substitution without SOC.
In our first-principles band structures of LaO$_{0.5}$F$_{0.5}$BiS$_2$, there is a small splitting less than 0.1 eV at the conduction band bottom on the X$'$ point for both two structures, which is not seen in LaOBiS$_2$ [see Fig.~\ref{fig:band_closeup}(a)(b)(d)(e)].
This is because the F substitution breaks the symmetry with respect to the layer exchange, which makes the electrostatic potentials on the two layers slightly different.
This issue, however, does not change which atomic orbitals are coupled or decoupled since it is determined by the mirror symmetries as shown in Table~\ref{table:orb}.
For example, at the X point for the tetragonal structure, all the Bi orbitals are still decoupled despite the small band splitting mentioned above.

Next, we shall analyze the effect of SOC on the band structure of LaOBiS$_2$.
The main role of SOC is to introduce the intrasite coupling between Bi $p_x$ and $p_y$ orbitals.
For the tetragonal structure, Hamiltonian at the X point where all the Bi orbitals are decoupled without SOC becomes two blocks by the intrasite SOC coupling: \{Bi$_1$ $p_x$, Bi$_1$ $p_y$, S$_2$ $p_x$, S$_2$ $p_y$\} and \{Bi$_2$ $p_x$, Bi$_2$ $p_y$, S$_1$ $p_x$, S$_1$ $p_y$\}.
The Bloch states now become a mixture of $p_x$ and $p_y$ orbitals, nevertheless, the degeneracy with respect to layer exchange still remains here. These situations are verified by our first-principles calculation as shown in Fig.~\ref{fig:band_closeup}(b)-(c). The change in the energy level of the conduction bands at the X point can be understood as a result of the orbital mixture.
Here, we find the strength of the intrasite SOC coupling on the Bi atom to be about 0.6 eV by constructing a tight-binding model of Bi $p_{x,y}$ and S $p_{x,y}$ with an inclusion of SOC in the same way presented before in this paper. Therefore, the strength of SOC is comparable to the energy difference between the Bi $p_x$ and $p_y$ bands at the X point for the tetragonal LaOBiS$_2$ without SOC, as shown in Fig.~\ref{fig:band_closeup}(b).

The situation regarding SOC is a bit complicated in the monoclinic structure.
As shown in Fig.~\ref{fig:band_closeup}(e), when SOC is not included, the first and third conduction bands from the bottom at the X point have a large weight of Bi $p_x$ orbitals, whereas the second and fourth bands have a large Bi $p_y$ weight.
On the other hand, the first and second conduction bands from the bottom at the X point have a large Bi $p_x$ weight when SOC is included as shown in Fig.~\ref{fig:band_closeup}(f). This change can be understood as a similar tendency to the tetragonal structure in the sense that SOC enlarges the splitting between the $p_x$ and $p_y$ bands.

\subsubsection{Superconductivity and bilayer effective models for its investigation}

Most theoretical studies of the superconductivity for the BiS$_2$-based superconductors in the tetragonal structure have been performed with a monolayer effective model~\cite{BiS2review3} because no band splitting is observed at the X point for their mother compounds.
On the other hand, our work reveals that an explicit treatment of two BiS$_2$ layers is necessary for the monoclinic structure owing to their large bilayer splitting at the X point.
For example, the monolayer model cannot describe the fact that a different amount of electron carriers reside in the bonding and antibonding bands that constitute much separated Fermi surfaces, as we have seen in Fig.~\ref{fig:FS}.
This situation where the bilayer coupling is of significant importance reminds us of a recent model construction study of $\beta$-ZrNCl, a superconductor with a bilayer honeycomb lattice structure, which revealed the presence of a surprisingly large bilayer coupling~\cite{Tanaka}.  This material is another superconductor with a relatively high $T_c$ of 15 K (HfNCl has $T_c$=25 K), whose pairing mechanism remains under debate~\cite{reviewMNX}. In fact, some commonalities between the BiS$_2$ superconductors and ZrNCl have been pointed out in an experimental study~\cite{Sakai}.

\begin{figure}
\begin{center}
 \includegraphics[width=8.5cm]{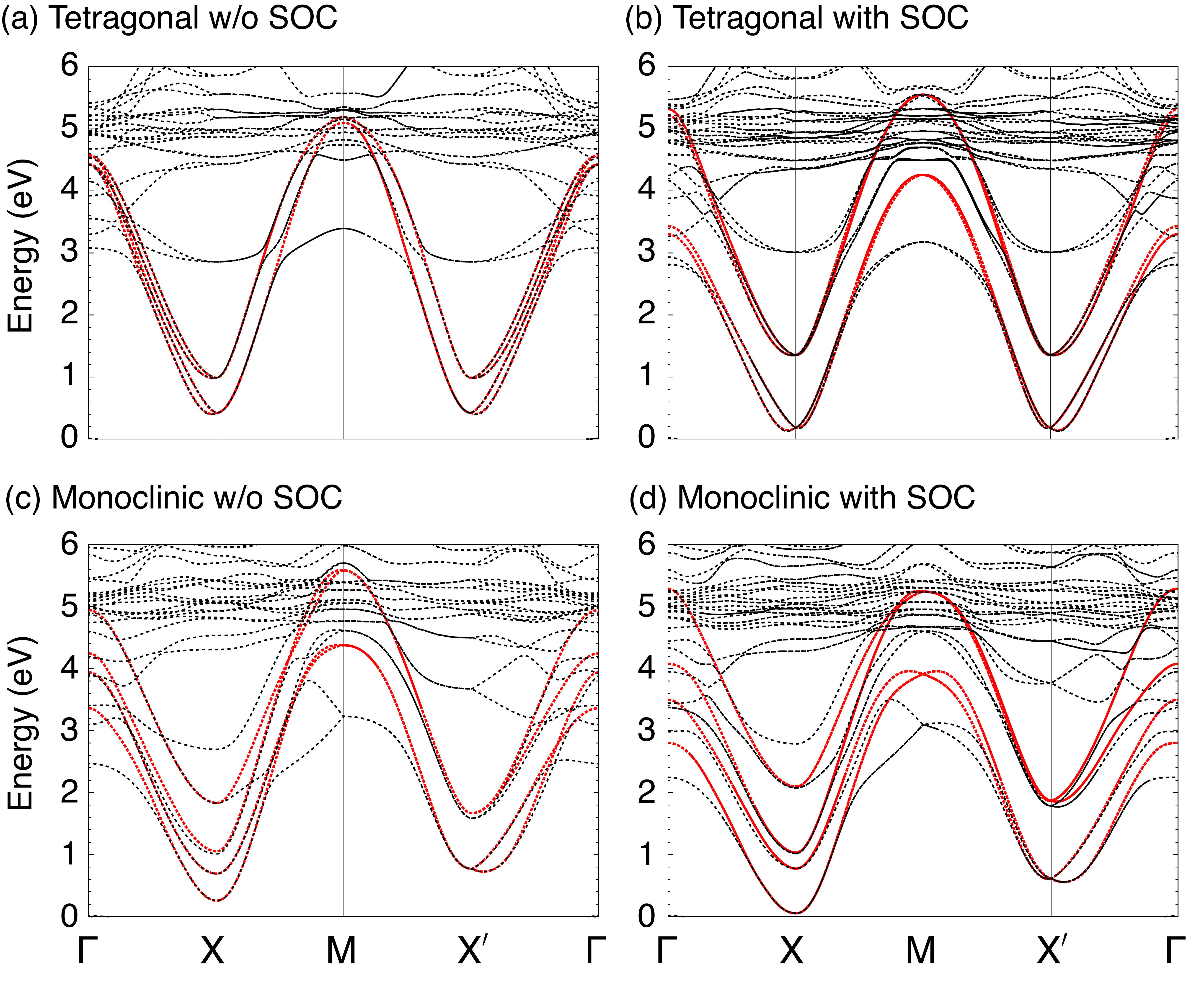}
 \caption{(Color online) Band structures of the bilayer Bi $p_{x,y}$ model (red solid lines) together with those obtained in first-principles calculations (black broken lines) for (a) the tetragonal structure without SOC, (b) that with SOC, (c) the monoclinic structure without SOC, and (d) that with SOC. The valence band top for each first-principles band structure is set to 0 eV.}
 \label{fig:Bimodel}
\end{center}
\end{figure}

Because the higher $T_c$ in the monoclinic structure will urge researchers to study the superconductivity using the bilayer effective model, we present parameters in the effective models consisting of Bi $p_{x,y}$ orbitals on two layers for the tetragonal and monoclinic structures of LaOBiS$_2$ (see Tables~\ref{table:param_woSOtetra}-\ref{table:param_wSOmono} and Fig.~\ref{fig:Bimodel}). Many computational details are the same as those presented in Sec.~3.2.1, but here the S sites are omitted from the effective models for convenience in theoretical studies, and hence the Wannier functions are the Bi-S hybridized orbitals. In addition, the effective models with and without SOC were constructed because SOC changes the order of the energy levels of the conduction bands at the X point for the monoclinic structure, as seen in Fig.~\ref{fig:band_closeup}(e)(f).
The spin moment of each Wannier function was set parallel to the $z$-axis.
For this purpose, we did not perform the maximal localization procedure in constructing the Wannier orbitals
because the localization process will mix the Wannier functions with different spin directions when SOC is included.
Maximal localization was not performed also for the models without SOC here so that one can compare the models with and without SOC. The frozen (inner) window was set to [0:2.0], [0:2.5], [0:1.0], and [0:0.8] eV for the tetragonal structure without SOC, that with SOC, the monoclinic structure without SOC, and that with SOC, respectively, where the valence top energy was set to 0 eV for each condition.
The outer window was set to [0:6.0] eV for all the models presented here.

We can find one important thing about the interlayer hopping amplitudes in the bilayer Bi $p_{x,y}$ effective models.
In Tables~\ref{table:param_woSOtetra} and \ref{table:param_woSOmono}, the maximum amplitude of the interlayer hopping process for the monoclinic structure (243 meV) is one order of magnitude larger than that in the tetragonal one (20 meV). 
This surprising difference is due to the shorter distance between Bi atoms in the monoclinic structure, as we have already addressed in Section 3.2.6.

\begin{table*}
\begin{center}
\begin{tabular}{c |c c c c}
\hline
\hline
 & \multicolumn{4}{c}{$\mu=$ 1} \\
$[\Delta x$, $\Delta y ]$ & $\nu=$ 1 &  2 & 3 & 4\\ \hline
$[-3,-1]$ & 16 (*) & - & -&-\\
$[-2,-3]$ & $-$11 (*) & - &- &-\\
$[-2,-2]$ & $-$14 (*) & $-$20 (*)& 11 (*) & 16 (*)\\
$[-2,-1]$ & 64 (*) & 28 (*) & -&-\\
$[-2,0]$ & $-$28 (*) & - &- &-\\
$[-1,-2]$ & 21 (*) & 28 & - &-\\
$[-1,-1]$ & 487 (*) & 389 (*) & $-$20 (*)& $-$15 (*)\\
$[-1,0]$ & $-$56 (*) & - & $-$20 & 15\\
$[0,-2]$ & 28 (*) & - & -&-\\
$[0,-1]$ & $-$280 (*) & - & $-$20 & 15\\
$[0,0]$ & 2833 (*) & - & $-$20 & $-$15\\
\hline
\hline
\end{tabular}
\end{center}
\caption{\label{table:param_woSOtetra} $t[\Delta x, \Delta y; \mu, \nu]$, hopping parameters from the $\nu$-th orbital in the $\mathbf{R}=\mathbf{0}$ unit cell to the $\mu$-th orbital in the $\mathbf{R}= (\Delta x, \Delta y, 0)$ unit cell, for the bilayer effective models of the tetragonal LaOBiS$_2$ without SOC presented in unit of meV. The unit cell is defined in Fig.~\ref{fig:model}(c) where only the Bi sites are taken here.
Orbital indices $\mu$ ($\nu$) $=1,2,3,$ and $4$ correspond to Bi$_1$ $p_{x}$, Bi$_1$ $p_y$, Bi$_2$ $p_x$, and Bi$_2$ $p_{y}$, respectively.
Values in and out of parentheses denote the model parameters for the monoclinic and tetragonal structures, respectively.
This table combined with the relations 
$t[\Delta x, \Delta y; \mu, \nu]=t[-\Delta x, -\Delta y; \tilde{\mu}, \tilde{\nu}]$ ($\tilde{\mu} = \mu+2$ ($\mu\leq2$), $\mu-2$ (otherwise); layer equivalency), 
$t[\Delta x, \Delta y; \mu, \nu]=(-1)^{\mu+\nu}t[-\Delta x-\delta x_{\mu}+\delta x_{\nu}, \Delta y;\mu, \nu]$ ($\delta x_{\mu}=1$ ($\mu\leq 2$), 0 (otherwise); M$_{\alpha}$ reflection, i.e., reflection with respect to the $yz$ plane where the Bi$_2$ site in the $\mathbf{R}=\mathbf{0}$ unit cell locates),
$t[\Delta x, \Delta y; \mu, \nu]=(-1)^{\mu+\nu}t[\Delta x, -\Delta y-\delta y_{\mu}+\delta y_{\nu};\mu, \nu]$ ($\delta y_{\mu}=1$ ($\mu\leq 2$), 0 (otherwise); M$_{\beta}$ reflection), 
$t[\Delta x, \Delta y; \mu, \nu]=(-1)^{\mu+\nu}t[-\Delta y-\delta y_{\mu}+\delta y_{\nu}, \Delta x; \mu ', \nu ']$ ($\mu ' =$ $\mu+1$ ($\mu$: odd), $\mu-1$ ($\mu$: even); $C_4$ rotation around the $z$ axis passing through the Bi$_2$ site in the $\mathbf{R}=\mathbf{0}$ unit cell),
and $t[\Delta x, \Delta y; \mu, \nu]=(t[-\Delta x, -\Delta y; \nu,\mu])^*$ (Hermiticity) covers all the hopping amplitudes larger than 10 meV for $\Delta z=0$.
Independent hopping parameters are marked as (*), from which the other ones can be generated.}
\end{table*}

\begin{table*}
\begin{center}
\begin{tabular}{c |c c c c c c}
\hline
\hline
 & \multicolumn{4}{c}{$\mu =$ 1} & \multicolumn{2}{|c}{2} \\
$[\Delta x$, $\Delta y ]$ & $\nu =$ 1 &  2 & 3 & 4 & \multicolumn{1}{|c}{2} & 4\\ \hline
$[-3,-2]$ & - & - & - & - & $-$10 (*) & - \\
$[-3, 0]$ & - & - & - & - & 16 (*) & -\\
$[-2,-3]$ &- & - & 10 (*) & - & - & -\\
$[-2,-2]$ & - & $-$42 (*) & 11 (*) & - & $-$19 (*)& -\\
$[-2,-1]$ & 44 (*) & 23 (*) & 19 (*) & - & 23 (*)& -\\
$[-2,0]$ & $-$55 (*) & - & 19 & - & 40 (*)& -\\
$[-1,-2]$ & 14 (*) & - & $-$23 (*) & $-$35 (*) & 49 (*)& $-$50 (*)\\
$[-1,-1]$ & 393 (*) & 296 (*) & $-$20 (*) & $-$33 (*) & 448 (*) & -\\
$[-1,0]$ & $-$53 (*) & - & $-$20 & 33 & $-$228 (*) & -\\
$[0,-3]$ & - & 17 (*) & - & - & - & -\\
$[0,-2]$ & $-$41 (*) & 62 (*) & 23 (*) & - & 22 (*) & 13 (*) \\
$[0,-1]$ & $-$310 (*) & 31 (*) & 140 (*) & $-$221 (*) & $-$99 (*) & 243 (*)\\
$[0,0]$ & 2887 (*) &- & 140 & 221 & 2915 (*) & 243\\
$[1,-3]$ & - & - & 11 (*) & - & - & -\\
$[1,-2]$ & 14 & $-$44 (*) & 53 (*) & $-$38 (*) & 49& -\\
$[1,-1]$ & 393 & $-$381 (*) & - & -& 448& 38 (*)\\
$[2,-2]$ & - & $-$23 (*) & - & -& $-$19 & -\\
$[2,-1]$ & 44 & $-$11 (*) & -& - & 23 & 11 (*)\\
\hline
\hline
\end{tabular}
\end{center}
\caption{\label{table:param_woSOmono} Hopping parameters (meV) for the bilayer effective models of the monoclinic LaOBiS$_2$ without SOC. Descriptions of the variables are the same as presented in the caption of Table~\ref{table:param_woSOtetra}.
This table combined with the relations 
$t[\Delta x, \Delta y; \mu, \nu]=t[-\Delta x, -\Delta y; \tilde{\mu}, \tilde{\nu}]$ ($\tilde{\mu} = \mu+2$ ($\mu\leq2$), $\mu-2$ (otherwise); layer equivalency), 
$t[\Delta x, \Delta y; \mu, \nu]=(-1)^{\mu+\nu}t[\Delta x, -\Delta y-\delta y_{\mu}+\delta y_{\nu};\mu, \nu]$ ($\delta y_{\mu}=1$ ($\mu\leq 2$), 0 (otherwise); M$_{\beta}$ reflection), 
and $t[\Delta x, \Delta y; \mu, \nu]=(t[-\Delta x, -\Delta y; \nu,\mu])^*$ (Hermiticity) covers all the hopping amplitudes larger than 10 meV for $\Delta z=0$.
Independent hopping parameters are marked as (*), from which the other ones can be generated.}
\end{table*}

\begin{table*}
\begin{center}
\begin{tabular}{c | c c c c c c c}
\hline
\hline
 & \multicolumn{7}{c}{($\mu$, $\sigma_{\mu})=$ (1,+1)} \\
$[\Delta x$, $\Delta y ]$ & ($\nu$, $\sigma_{\nu})=$ (1,+1) & (1,$-$1) & (2,+1) & (2,$-$1) & (3,+1) & (4,+1) & (4,$-$1) \\ \hline
$[-3,-1]$ & 16 (*) & -& -& -& -& -& -\\
$[-2,-3]$ & $-$10 (*)& -& -& -& -& -& -\\
$[-2,-2]$ & $-$16 (*) & -& $-$26$-$4$i$ (*) & - & 10 (*) & 16+1$i$ (*)& -\\
$[-2,-1]$ & 55 (*) & -& 26$-$5$i$ (*) & - & - & -&-\\
$[-2,0]$ & $-$27 (*) & -& -& - & -& -& -\\
$[-1,-2]$ & 25 (*) & -& 26$-$5$i$ & - & -& -& -\\
$[-1,-1]$ & 454 (*) & $-$22$-$1$i$ (*) & 367$-$36$i$ (*)& $-$12+12$i$ (*) & $-$19 (*) &$-$19$-$3$i$ (*) & 18+18$i$ (*)\\
$[-1, 0]$ & $-$57 (*) & -& $-$19$i$ (*) & -& $-$19 & 19$-$3$i$ & $-$18+18$i$ \\
$[0,-3]$ & 11 (*) & - &- & - &- & - &-\\
$[0,-2]$ & 24 (*) & -& - & -&- & -& -\\
$[0,-1]$ & $-$258 (*) & - & $-$19$i$ & - & $-$19 & 19$-$3$i$ & 18$-$18$i$\\
$[0,0]$ & 2771 (*) & - & $-$598i (*) & - & $-$19 & $-$19$-$3$i$ & $-$18$-$18$i$ \\
\hline
\hline
\end{tabular}
\end{center}
\caption{\label{table:param_wSOtetra} Hopping parameters (meV) for the bilayer effective models of the tetragonal LaOBiS$_2$ with SOC. Descriptions of the variables are the same as presented in the caption of Table~\ref{table:param_woSOtetra} except that the spin index $\sigma_{\mu}=\pm 1$ (up or down) for the $\mu$-th orbital is introduced here.
This table combined with the relations $t[\Delta x, \Delta y; (\mu, \sigma_{\mu}), (\nu, \sigma_{\nu})]=\sigma_{\mu}\sigma_{\nu}(t[\Delta x, \Delta y; (\mu, -\sigma_{\mu}), (\nu, -\sigma_{\nu})])^*$ (time-reversal symmetry), 
$t[\Delta x, \Delta y; (\mu, \sigma_{\mu}), (\nu, \sigma_{\nu})]=t[-\Delta x, -\Delta y; (\tilde{\mu}, \sigma_{\mu}), (\tilde{\nu}, \sigma_{\nu})]$ ($\tilde{\mu} = \mu+2$ ($\mu\leq2$), $\mu-2$ (otherwise); layer equivalency), 
$t[\Delta x, \Delta y; (\mu, \sigma_{\mu}), (\nu,\sigma_{\nu})]=(-1)^{\mu+\nu}t[-\Delta x-\delta x_{\mu}+\delta x_{\nu}, \Delta y; (\mu,-\sigma_{\mu}), (\nu,-\sigma_{\nu})]$ ($\delta x_{\mu}=1$ ($\mu\leq 2$), 0 (otherwise); M$_{\alpha}$ reflection), 
$t[\Delta x, \Delta y; (\mu, \sigma_{\mu}), (\nu,\sigma_{\nu})]=(-1)^{\mu+\nu}\sigma_{\mu}\sigma_{\nu}t[\Delta x, -\Delta y-\delta y_{\mu}+\delta y_{\nu}; (\mu,-\sigma_{\mu}), (\nu,-\sigma_{\nu})]$ ($\delta y_{\mu}=1$ ($\mu\leq 2$), 0 (otherwise); M$_{\beta}$ reflection), 
$t[\Delta x, \Delta y; (\mu, \sigma_{\mu}), (\nu, \sigma_{\nu})]=(-1)^{\mu+\nu}\mathrm{i}^{(-\sigma_{\mu}+\sigma_{\nu})/2}t[-\Delta y-\delta y_{\mu} +\delta y_{\nu}, \Delta x; (\mu ', \sigma_{\mu}), (\nu ', \sigma_{\nu})]$ ($\mu ' =$ $\mu+1$ ($\mu$: odd), $\mu-1$ ($\mu$: even); C$_4$ rotation),
and $t[\Delta x, \Delta y; (\mu, \sigma_{\mu}), (\nu, \sigma_{\nu})]=(t[-\Delta x, -\Delta y; (\nu, \sigma_{\nu}), (\mu, \sigma_{\mu})])^*$ (Hermiticity), covers all the hopping amplitudes larger than 10 meV for $\Delta z=0$.
Independent hopping parameters are marked as (*), from which the other ones can be generated.}
\end{table*}

\begin{table*}
\begin{center}
\begin{tabular}{c |c c c c c c c c c c}
\hline
\hline
 & \multicolumn{7}{c}{($\mu$, $\sigma_{\mu})=$ (1,+1)} & \multicolumn{3}{|c}{(2,+1)} \\
$[\Delta x$, $\Delta y ]$ & $(\nu$, $\sigma_{\nu})=$ (1,+1) & (1,$-$1) & (2,+1) & (2,$-$1) & (3,+1) & (4,+1) & (4,$-$1) & \multicolumn{1}{|c}{(2,+1)}  & (2,$-$1) & (4,+1)\\ \hline
$[-3,-2]$ & -& -& - & -& -
& -&- & $-$10$-$2$i$ (*)& -& -\\
$[-3,0]$ & -& -& - & -& -
& -&- & 22 (*)& -& -\\
$[-2,-3]$ & -& -& -& -& -
& - & 4$-$14$i$ (*)& - & - & - \\
$[-2,-2]$ & -& -& $-$30$-$5$i$ (*)& 9+20$i$ (*)& 11 (*)
& 6$-$11$i$ (*) & 1$-$17$i$ (*)& $-$34$-$3$i$ (*) & -& -\\
$[-2,-1]$ & 36+2$i$ (*)& -& 12+1$i$ (*)& $-$18$-$10$i$ (*)&  25 (*)
& 1+11$i$ (*) & $-$8+10$i$ (*)& 34$-$2$i$ (*)& -& -\\
$[-2,0]$ & $-$79 (*)& -& $-$10$i$ (*) & $-$14$i$ (*) & 25
& $-$1+11$i$ & 8+10$i$ & 49 (*) & $-$10 (*) &- \\
$[-1,3]$ & -& -& - & -& -
& -&- & 10+3$i$ (*) & -& -\\
$[-1,-2]$ & 13$-$1$i$ (*)& -&10+8$i$ (*)& -& $-$18 (*)
& $-$22+17$i$ (*)& 34$-$18$i$ (*)& 52$-$9$i$ (*)& 3$-$11$i$ (*)& $-$43 (*)\\
$[-1,-1]$ & 356$-$20$i$ (*)& 48$-$14$i$ (*)& 205+20$i$ (*)& $-$222$-$39$i$ (*) & -
& $-$24$-$45$i$ (*)& 19+34$i$ (*)& 417$-$5$i$ (*)& 6$-$42$i$ (*)& $-$43\\
$[-1,0]$ & $-$47 (*)& $-$11 (*) & -& $-$30$i$ (*)& -
& 24$-$45$i$  & $-$19+34$i$ & $-$203 (*)&-&- \\
$[0,-3]$ & -& -& 11$-$1$i$ (*)& $-$5$-$11$i$ (*)& -
& $-$17$-$3$i$ (*)&  $-$12+6$i$ (*)& 14$-$1$i$ (*)& -&- \\
$[0,-2]$ & $-$64+19$i$ (*)& - & 30$-$4$i$ (*)& $-$20$-$27$i$ (*) & 12 (*)
& 15$-$10$i$ (*)& $-$15$-$2$i$ (*) & 34+9$i$ (*)& -&15 (*)\\
$[0,-1]$ & $-$294+13$i$ (*)&- & 29$-$27$i$ (*) & $-$6+36$i$ (*)& 123 (*)
& $-$129$-$31$i$ (*)& 132+10$i$ (*)& $-$81$-$9$i$ (*)& - & 227 (*)\\
$[0,0]$ & 2801 (*)& -&  $-$364$i$ (*) & $-$224$i$ (*)&123
&129$-$31$i$ & $-$132+10$i$ & 2847 (*)& -& 227\\
$[1,-3]$ &- &- & $-$13$-$4$i$ (*)& -& -
&- &- &10+3$i$ &- &- \\
$[1,-2]$ &13$-$1$i$ & -& $-$47$-$3$i$ (*)& 26$-$8$i$ (*) & 41 (*)
& $-$25$-$16$i$ (*)& 10+27$i$ (*)& 52$-$9$i$ & $-$3$-$11$i$ &$-$11 (*)\\
$[1,-1]$ & 356$-$20$i$ & $-$48$-$14$i$ & $-$230$-$19$i$ (*)& 293$-$18$i$ (*) & $-$19 (*)
& $-$17+15$i$ (*)& 17$-$20$i$ (*)& 417$-$5$i$ & $-$6$-$42$i$ & 52 (*)\\
$[1,0]$ & $-$47 & 11 & -& $-$23$i$ (*) & $-$19
& 17+15i &$-$17$-$20$i$ & $-$203 & -& 52\\
$[2,-3]$ & -& -& -& 7+9$i$ (*) & -
& - & 11$-$1$i$ (*)& -& -&- \\
$[2,-2]$ & -& -& 10$-$1$i$ (*)& 9+28$i$ (*)& -
& 1+11$i$ (*) & -& $-$34$-$3$i$ &- & -\\
$[2,-1]$ & 36+2$i$&  -& - & 16$-$15$i$ (*) & $-$12 (*)
& $-$15$-$9$i$ (*)& -& 34$-$2$i$ &- & 17 (*)\\
$[2,0]$ & $-$79 & -& -& $-$39$i$ (*)& $-$12
&15$-$9$i$ &- & 49 & 10 & 17\\
\hline
\hline
\end{tabular}
\end{center}
\caption{\label{table:param_wSOmono} Hopping parameters (meV) for the bilayer effective models for the monoclinic LaOBiS$_2$ with SOC. Descriptions of the variables are the same as those presented in the caption of Table~\ref{table:param_wSOtetra}.
This table combined with the relations $t[\Delta x, \Delta y; (\mu, \sigma_{\mu}), (\nu, \sigma_{\nu})]=\sigma_{\mu}\sigma_{\nu}(t[\Delta x, \Delta y; (\mu, -\sigma_{\mu}), (\nu, -\sigma_{\nu})])^*$ (time-reversal symmetry), 
$t[\Delta x, \Delta y; (\mu, \sigma_{\mu}), (\nu, \sigma_{\nu})]=t[-\Delta x, -\Delta y; (\tilde{\mu}, \sigma_{\mu}), (\tilde{\nu}, \sigma_{\nu})]$ ($\tilde{\mu} = \mu+2$ ($\mu\leq2$), $\mu-2$ (otherwise); layer equivalency), 
$t[\Delta x, \Delta y; (\mu, \sigma_{\mu}), (\nu,\sigma_{\nu})]=(-1)^{\mu+\nu}\sigma_{\mu}\sigma_{\nu}t[\Delta x, -\Delta y-\delta y_{\mu}+\delta y_{\nu}; (\mu,-\sigma_{\mu}), (\nu,-\sigma_{\nu})]$ ($\delta y_{\mu}=1$ ($\mu\leq 2$), 0 (otherwise); M$_{\beta}$ reflection), 
and $t[\Delta x, \Delta y; (\mu, \sigma_{\mu}), (\nu, \sigma_{\nu})]=(t[-\Delta x, -\Delta y; (\nu, \sigma_{\nu}), (\mu, \sigma_{\mu})])^*$ (Hermiticity), covers all the hopping amplitudes larger than 10 meV for $\Delta z=0$. For example, $t[1, 1; (1, +1), (2, +1)]$ has fifteen equivalent hopping paths that can be generated by applying the above relations.
Independent hopping parameters are marked as (*), from which the other ones can be generated.}
\end{table*}

\subsubsection{Relationship between our study and some other studies}

Our observation of the orbital characters of Bloch states in the lowest conduction band [Fig.~\ref{fig:orbitals}(b)-(d)] for the tetragonal structure is consistent with the recent ARPES experiment that shows that the contribution of the Bi $p_x$ orbitals is dominant along the $\Gamma$-X line, whereas that of the Bi $p_y$ orbitals appears along the X-M line on the Fermi surface of CeO$_{0.5}$F$_{0.5}$BiS$_2$~\cite{Sugimoto_px}.

Our work focuses on a large bilayer coupling induced by the symmetry breaking of the crystal structure. The relationship between the symmetry of the layered systems and interlayer coupling has recently been generalized by one of the authors~\cite{LayerShift}. One interesting example is the post-graphene material MoS$_2$, where the valley excitonic state exhibits anomalous two-dimensionality~\cite{3RMoS2}.

\section{Conclusions\label{5}}

We have performed first-principles band structure calculations on the tetragonal and monoclinic structures of LaO$_{0.5}$F$_{0.5}$BiS$_2$ and have found some important differences between them. The monoclinic band structure exhibits a sizable band splitting, e.g., at the conduction band bottom of the X point, which induces a substantial change of the Fermi surface topology. The origin of such splitting is the strong bilayer coupling induced by the symmetry breaking of the crystal structure, which is clearly shown by our analysis using the tight-binding model. Anisotropy with respect to the $x$- and $y$-directions is also an important feature of the monoclinic structure. 
Because of its higher $T_c$, further investigation on the monoclinic structure of the BiS$_2$-based superconductors using the knowledge obtained in our study is expected.

\section*{Acknowledgments}
This study was supported by Grant-in-Aid for Young Scientists (B) (Nos. JP15K17724 and JP15K20940) and Grant-in-Aid for Scientific Research (A) (No. JP26247057) from the Japan Society for the Promotion of Science.


\begin{thebibliography}{999}
\bibitem{foot:pXY} Our basis functions are different from those used in the previous study~\cite{Usui} that used the $p_{X,Y}$ orbitals, which are rotated by 45 degrees from the $p_{x,y}$ orbitals. We adopt the $p_{x,y}$ orbitals as basis functions in our model because they are much suited to the lowered crystal symmetry in the monoclinic structure where the reflection symmetry with respect to the M$_{\alpha}$ plane is broken.
\end{thebibliography}
\end{document}